\documentclass[conference,tightenlines]{IEEEtran}
\usepackage{graphicx}
\usepackage{amssymb}
\usepackage{epsfig}
\usepackage{hyperref}
\usepackage{fancyhdr}

\usepackage{soul}  

\usepackage[usenames, dvipsnames]{color}
\usepackage{algorithmic}
\usepackage{algorithm}
\usepackage[tight,footnotesize]{subfigure}
\begin{document}

\newcommand\andre[1]{{\color{blue}[ANDR\'{E}: #1]}}
\newcommand\enrico[1]{{\color{blue}[ENRICO: #1]}}
\newcommand\evan[1]{{\color{magenta}[EVAN: #1]}}
\newcommand\amy[1]{{\color{ForestGreen}[AMY: #1]}}
\newcommand\arjun[1]{{\color{cyan}[ARJUN: #1]}}
\newcommand\ken[1]{{\color{red}[KEN: #1]}}

\newcommand\Fig[1]{Fig.~\ref{fig:#1}}
\newcommand\eq[1]{Eq.~\ref{eq:#1}}
\newcommand\sect[1]{Sect.~\ref{sec:#1}}
\newcommand\refcite[1]{Ref.~\cite{#1}}
\newcommand\refscite[1]{Refs.~\cite{#1}}
%
\title{Simulating the \textit{weak} death of the neutron in a femtoscale universe with near-Exascale computing}



%
\author{
\IEEEauthorblockN{
Evan Berkowitz\IEEEauthorrefmark{1},
M.A. Clark\IEEEauthorrefmark{2},
Arjun Gambhir\IEEEauthorrefmark{3}\IEEEauthorrefmark{4}\IEEEauthorrefmark{5},
Ken McElvain\IEEEauthorrefmark{5}\IEEEauthorrefmark{4},
Amy Nicholson\IEEEauthorrefmark{6},\\
Enrico Rinaldi\IEEEauthorrefmark{7}\IEEEauthorrefmark{4},
Pavlos Vranas \IEEEauthorrefmark{3}\IEEEauthorrefmark{4}
Andr\'{e} Walker-Loud \IEEEauthorrefmark{4}\IEEEauthorrefmark{3}\IEEEauthorrefmark{5},\\
and\ Chia Cheng Chang\IEEEauthorrefmark{4},
B\'{a}lint Jo\'{o}\IEEEauthorrefmark{8},
Thorsten Kurth\IEEEauthorrefmark{9}\IEEEauthorrefmark{4},
Kostas Orginos\IEEEauthorrefmark{10}\IEEEauthorrefmark{11},
}
\smallskip
\IEEEauthorblockA{\IEEEauthorrefmark{1}
	Institut f\"{u}r Kernphysik and Institute for Advanced Simulation,
	Forschungszentrum J\"{u}lich,
	54245 J\"{u}lich Germany}
\IEEEauthorblockA{\IEEEauthorrefmark{2}
	NVIDIA Corporation,
	Santa Clara, California, 95051, USA}
\IEEEauthorblockA{\IEEEauthorrefmark{3}
	Physical Sciences Directorate,
	Lawrence Livermore National Laboratory,
	Livermore, California 94550, USA}
\IEEEauthorblockA{\IEEEauthorrefmark{4}
	Nuclear Science Division,
	Lawrence Berkeley National Laboratory,
	Berkeley, CA 94720, USA}
\IEEEauthorblockA{\IEEEauthorrefmark{5}
	Department of Physics,
	University of California,
	Berkeley, CA 94720, USA}
\IEEEauthorblockA{\IEEEauthorrefmark{6}
	Department of Physics and Astronomy,
	University of North Carolina,
	Chapel Hill, NC 27516-3255, USA}
\IEEEauthorblockA{\IEEEauthorrefmark{7}
	RIKEN-BNL Research Center,
	Brookhaven National Laboratory,
	Upton, NY 11973, USA}
\IEEEauthorblockA{\IEEEauthorrefmark{8}
	Scientific Computing Group,
	Thomas Jefferson National Accelerator Facility,
	Newport News, VA 23606, USA}
\IEEEauthorblockA{\IEEEauthorrefmark{9}
	NERSC,
	Lawrence Berkeley National Laboratory,
	Berkeley, CA 94720, USA}
\IEEEauthorblockA{\IEEEauthorrefmark{10}
	Department of Physics,
	The College of William \& Mary,
	Williamsburg, VA 23187, USA}
\IEEEauthorblockA{\IEEEauthorrefmark{11}
	Theory Center,
	Thomas Jefferson National Accelerator Facility,
	Newport News, VA 23606, USA}
}

\maketitle

\begin{abstract} 

  The fundamental particle theory called Quantum Chromodynamics (QCD)
  dictates everything about protons and neutrons, from their intrinsic
  properties to interactions that bind them into atomic nuclei.
  Quantities that cannot be fully resolved through experiment, such as
  the neutron lifetime (whose precise value is important for the
  existence of light-atomic elements that make the sun shine and life
  possible), may be understood through numerical solutions to QCD.  We
  directly solve QCD using Lattice Gauge Theory and calculate nuclear
  observables such as neutron lifetime.  We have developed an improved
  algorithm that exponentially decreases the time-to-solution and
  applied it on the new CORAL supercomputers, Sierra and Summit.  We
  use run-time autotuning to distribute GPU resources, achieving 20\%
  performance at low node count.  We also developed optimal
  application mapping through a job manager, which allows CPU and GPU
  jobs to be interleaved, yielding 15\% of peak performance when
  deployed across large fractions of CORAL.

\end{abstract}

\section{\label{sec:just} Justification for ACM Gordon Bell Prize}
Our algorithm {\it exponentially decreases time to solution} by accessing parameter space where the signal-to-noise is exponentially larger.
Our code is optimally mapped on the machine and our kernel can ``auto-tune'' local GPU resources at runtime.
We achieve excellent scaling across large fractions of the CORAL
supercomputers achieving a peak sustained performance in excess of 15\%.%
\footnote{This work has been submitted to the IEEE for possible publication. Copyright may be transferred without notice, after which this version may no longer be accessible.}

\section{\label{sec:perf_attributes} Performance Attributes}

\begin{table}[h]
\centering
\begin{tabular}{|c|c|}
\hline
Attribute                    & Value \\
\hline
Category of achievement & time to solution \\
\hline
method                    & explicit \\
\hline
reporting                   & whole application including I/O \\
\hline
precision                   & mixed-precision \\
\hline
system scale              & full-scale system \\
\hline
measurement method & FLOP count \\
\hline
\end{tabular}
\caption{\label{tab:attributes} Performance Attributes}
\end{table}

\section{\label{sec:overview} Overview }

The Standard Model (SM) of particle physics is the currently accepted (and incredibly well-tested) theory describing fundamental particles and their interactions, which give rise to all ordinary matter in the universe. 
Calculations of properties of matter directly from the SM can give us unique insight into some of the outstanding questions about our universe that are difficult or impossible to address with experiment. 
For example, when compared to all other atomic nuclei, the simplest bound nucleus, Deuterium (one proton and one neutron), has an anomalously small binding energy per constituent.
This small energy has huge consequences for the evolution of life as we know it, in part from its contribution to the very slow fusion process which has heated our Sun for the last 4.6 billion years and prevents it from collapsing under its own gravity.
The question, then, is whether this anomalous value is a chance consequence of fine-tuning of the parameters of the SM, or a more robust, fundamental property of the theory; only direct calculations from the SM may provide an answer.

The lifetime of the neutron is an equally important quantity for the evolution of the universe as we know it. 
Free neutrons end their lives by decaying into a proton, electron, and anti-neutrino within an average time of about 14 minutes.
In contrast, when bound in nuclei, such as Deuterium, or Helium (He) (2 protons and 1 or 2 neutrons), neutrons become stable, and their lifetime, infinite.
This stability has importance consequences: in the first few minutes after the Big Bang, as the universe was expanding and cooling, free neutrons decayed, with the resulting protons later binding with electrons to form Hydrogen atoms.
Some neutrons were saved from this decay by being bound into light nuclei, such that the primordial universe was composed of approximately 75\% Hydrogen (H) and 25\% ${}^4$He by mass fraction, or rather, there was approximately 1 neutron for every 7 protons.
Had the neutron lifetime differed from this value, H and He would have formed at different rates and the universe today would be quite different. For example, this initial abundance of hydrogen is critical to the formation of stars, including our Sun.

The neutron lifetime can be precisely measured experimentally.
However, current results coming from two different measurement techniques (trapped neutrons versus neutron beams) disagree significantly: $\tau_n^{trap}=879.4(6)s$ and $\tau_n^{beam}=888(2)s$~\cite{Patrignani:2016xqp, Czarnecki:2018okw}.
This has raised the exciting possibility that there may be new, unknown physics contributing to the neutron decays which gives rise to this difference~\cite{Fornal:2018eol,Tang:2018eln,Serebrov:2018mva}.

To decide whether new physics causes this discrepancy, we must first confidently know the neutron lifetime as predicted by all known physics and then compare it to experimental results.
The SM of particle physics makes a definite prediction (recently updated in Ref.~\cite{Czarnecki:2018okw}), a lifetime of
\begin{equation}\label{eq:t_n}
    \tau_n = \frac{5172.0 \pm 1.0 }{1 + 3 g_A^2}\, \textrm{ seconds.}
\end{equation}
The \emph{axial coupling of the neutron}, $g_A$, dictates the strength of the \emph{weak} interaction with protons and neutrons, and has implications beyond the neutron lifetime. For example, it is also a key quantity in determining the rate of solar fusion as it dictates the rate of conversion of the protons in H to neutrons at high temperatures during the formation of deuterium.

The axial coupling is not an input parameter of the SM but an emergent property of neutrons, themselves made of \emph{quarks} and \emph{gluons} which are described by a part of the Standard Model known as Quantum Chromodynamics (QCD).
Computing $g_A$ from QCD requires high performance computing, as it entails solving the equations non-perturbatively.
QCD is so \emph{strongly interacting} that we never see isolated quarks or gluons---we only see them bound into protons, neutrons, and other \emph{hadrons}.
QCD fixes the nuclear interactions between protons and neutrons, which are strong enough to bind them into nuclei, overcoming the protons' electromagnetic repulsion.

That these interactions are so strong means that we cannot rely on pen-and-paper estimates from QCD but must resort to numerical methods instead.
\emph{Lattice QCD} (LQCD) is the formulation of QCD on a finite space-time grid, or lattice, representing a portion of our universe, which captures the correct quantum-mechanical behavior via a Monte Carlo approach.
While the theory and methods are known, leveraging this approach for nuclear physics proves to be extremely demanding computationally.
For example, the LQCD community previously estimated that a determination of $g_A$ with a 2\% precision would be possible by approximately 2020 with the CORAL generation of supercomputers~\cite{usqcd_doe_2016}.
In order to have an impact on the experimental discrepancy in the measurement of the neutron lifetime, we require a determination of $g_A$ with a precision of 0.2\% or better.

While pursuing the challenging calculation of $g_A$, we have developed a new physics algorithm~\cite{Bouchard:2016heu} which allowed for exponential improvement over previous state-of-the-art results.
We have developed new software to exploit this idea and performed a full-scale calculation of $g_A$ using recent generations of NVIDIA-GPU accelerated compute nodes, with newly developed job management software to achieve a determination of $g_A$ with an unprecedented 1\% precision~\cite{Berkowitz:2017gql,Chang:2017oll}. 

This benchmark calculation showed for the first time that lattice QCD may be reliably applied to nuclear physics applications with fully controlled uncertainties, opening the door to applications to light nuclei.
Even in this preliminary stage, the work was important enough to merit publication in Nature~\cite{Chang:2018uxx}.
Furthermore, we have critically identified how increased calculation time can systematically and simultaneously improve the three dominant sources of uncertainty in the calculation of $g_A$.

In order to complete the calculation of the neutron lifetime to the precision necessary for distinguishing new physics, as well as broaden the successful application of lattice QCD to light nuclei, we must apply our techniques to multi-nucleon systems, which requires us to calculate on larger lattices, increasing their size to unprecedented proportions. With these large lattices in hand, direct calculation of nuclear properties are unlocked.  Coupled with other theoretical tools, first-principles determinations may reach as deep into the periodic table as oxygen.

\section{\label{sec:curent_sa} Current State of the Art}

\noindent{\bf Background}

Solving the equations of QCD using numerical LQCD methods rely on fast
linear solves of a very large sparse matrix whose size is determined
by the size of the four-dimensional lattice used to discretize the
equations (and including a fifth dimension in state-of-the-art
computations).  On a lattice of spatial size $L$ and temporal size $T$
(with extra dimension $L_5$) the matrix, called the Dirac operator,
acts on a space of size $L^3 \times T \times L_5 \times N_c \times
N_s$, where $N_c=3$ is the size of the fundamental representation of \(SU(3)\) (which describes the
quark-gluon interactions), and $N_s=4$ counts the quark spin degrees of freedom.  Typical lattices are
$48^3 \times 64 \times 20$ and $64^3\times96\times12$, but understanding the properties of light
nuclei will require bigger ones (for example a
$96^3 \times 144 \times 20$ lattice, whose Dirac operator has
$\sim 10^{19}$ entries).  The discretized Dirac operator is radius-one
stencil kernel, and contains (complex-valued) ``submatrices'' along
the diagonal that are small, $12 \times 12$, but dense.  As is typical
with stencil computations, the usual prescription is to
\begin{enumerate}
\item Pack the halo into contiguous buffers
\item Communicate halos to neighbors
\item Compute the interior stencil application
\item Once halos have been communicated complete halo stencil computation
\end{enumerate}
In an ideal world the communication (2) can be completely overlapped with the interior GPU computation (3), and when this is achieved the computation will successfully strong scale.  

Much of the computational cost for LQCD goes into using iterative
Krylov-solver methods to solve the Dirac operator and obtain the
so-called ``propagator''.  For the discretization used in the present
work, the so-called M\"{o}bius Domain-Wall discretization, the
state-of-the art technique is to utilize conjugate gradient on the
normal equations. This is the first challenge and it can be overcome by using GPU-accelerated nodes on the CORAL systems and by scaling to a larger number of nodes when the lattice size increases.
By increasing the number of calculated propagators we can reduce the statistical uncertainty of the final observable, because each new propagator amounts to a new sample in the ensemble describing how quarks propagate in the lattice universe.
However, the statistical uncertainty is only reduced $\propto N_{sample}^{-1/2}$.

\noindent {\bf Performance related}
For comparison of our code performance with the previous state of the
art we use metrics that we have measured on the previous generation
supercomputer, Titan.
The results are shown in figure \ref{fig:strongscale_subgroup}.

We utilize the Chroma~\cite{Edwards:2004sx} application in our work:
Chroma is part of the SciDAC-funded USQCD software suite~\cite{usqcd}
that achieves portability through utilizing optimized libraries for
the time-consuming portions of the QCD, e.g., the Dirac linear system
solves.  In the case of NVIDIA GPUs, QUDA is used~\cite{Clark:2009wm}
to allow Chroma to offload onto the GPUs.  QUDA is a framework for
constructing highly-optimized QCD algorithms, written using CUDA C++,
with specific focus on linear-solver algorithms.  High performance is
achieved using techniques such as kernel fusion to minimize memory
traffic, matrix-free stencil kernels, and kernel autotuning to
maximize performance.  Performance is maximized using mixed-precision
iterative solvers, with the optimum approach for the stencil at hand
being to use a red-black preconditioned {\it double-half} CG solver,
where most of the work is done using 16-bit precision fixed-point
storage (utilizing single-precision computation) with occasional
reliable updates to full double precision~\cite{Clark:2009wm}.

\noindent{\bf GPU Kernel Autotuning}

GPU kernel autotuning is important in order to maximize cache hit rate
as well as ensure performance portability across GPU generation.  The
QUDA autotuner is a C++ run-time autotuner, where a brute-force search
through launch parameter space is performed the first time an un-tuned
kernel or algorithm is encountered.  Once the optimum launch
configuration is known, this is stored in a \texttt{std::map}, and is
subsequently looked up on demand if and when this kernel is
encountered again.  Each entry in the map is given a unique identifier
which stores the optimum launch parameters, as well as other metadata,
such as performance metrics for the given kernel.  The class structure
makes it easy to manage the backup/restore of input data in the case
of data-destructive algorithms, and to specify unique look-up keys for
the same kernel but with slightly different parameters to ensure that
a given computation is always launched with the optimum parameters.

\noindent {\bf Physics algorithms related}

\begin{figure}[!t]
  \centering
  {\includegraphics[width=.90\columnwidth,angle=0]{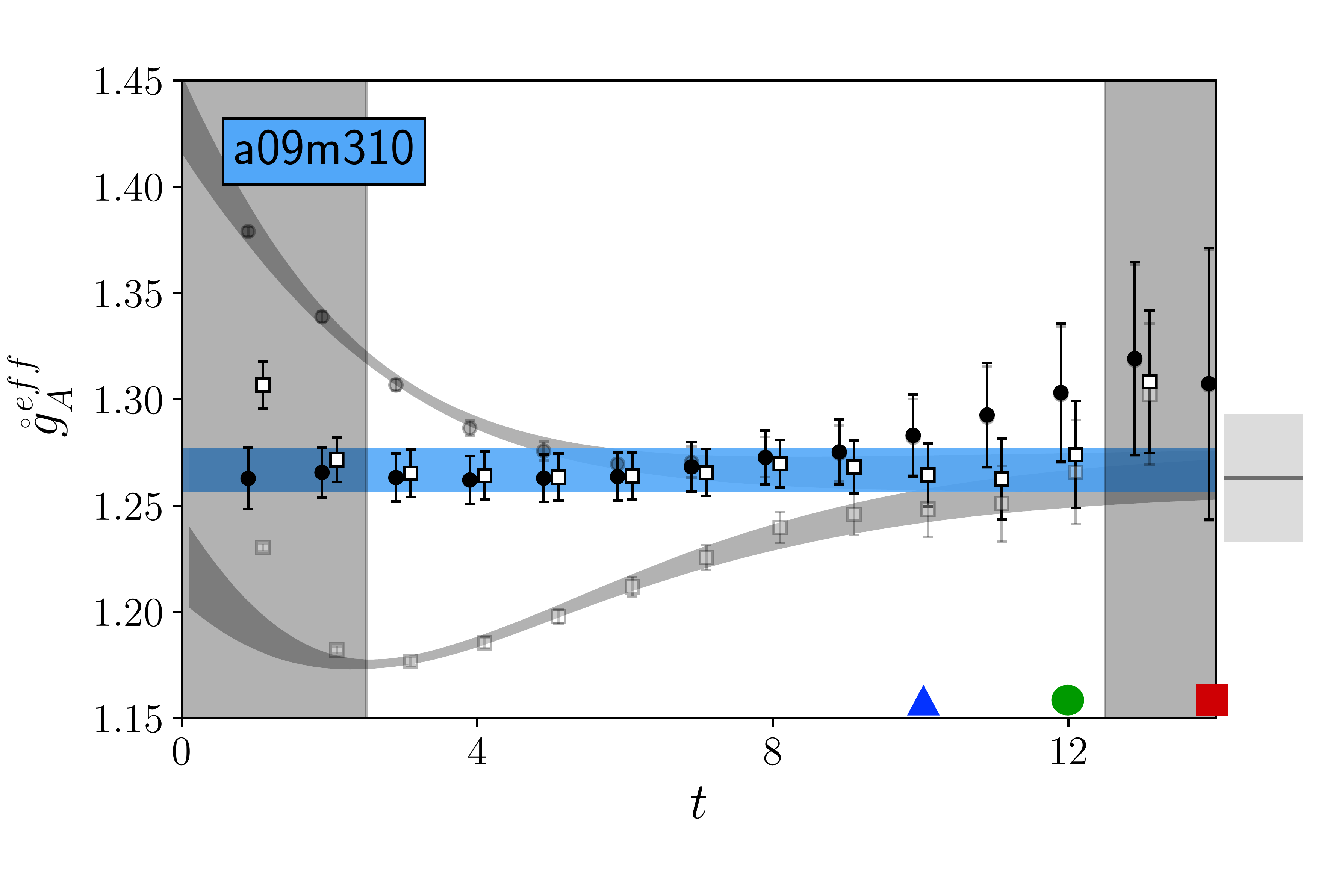}}
  \caption{Our modified results are shown in grey together with a fit to extract $g_A$ in the large $t$ region. The results after removing the contribution from excited neutron modelled by the fit are shown in black and white symbols. The signal is extracted from points at small temporal distances. On the contrary, traditional calculations rely on time separations indicated by the colored symbols (triangle, circle and square) at large $t$ where the noise is exponentially larger. The blue band is our final answer, while the grey band on the right of the plot is the one obtained from a traditional calculation, which was obtained with an order of magnitude larger statistical sample.}
  \label{fig:ga_eff}
\end{figure}

Traditional methods use a collection of propagators tied together to extract the neutron axial coupling $g_A$, by looking at its behavior for large temporal distances on the lattice. Unfortunately, the signal-to-noise ratio decreases exponentially at those large separations and shorter separations are typically not used because the signal there is affected by excited neutron effects which can systematically shift the value of $g_A$. It seems hopeless to try and beat an exponential decrease in statistical signal with a power law increase in samples. Therefore, the second challenge is to find new techniques yielding a final answer with an uncertainty significantly smaller by exploiting shorter temporal separations, currently missed by state-of-the-art calculations. For that reason we designed a new type of propagator~\cite{Bouchard:2016heu} which yields all the temporal distances for the cost of one temporal distance in the traditional method.
Using this early-time data we can determine and subtract away the systematic contamination and extract meaningful physics from very precise but otherwise uninformative data.
An example of $g_A$ obtained from our new algorithm is shown in Fig.~\ref{fig:ga_eff}. To compare with traditional LQCD calculation we notice that we can obtain a remarkably more precise answer with an order of magnitude smaller number of statistical samples~\cite{Chang:2017oll}.

\section{\label{sec:innovations} Innovations}

\noindent{\bf Dense-node Optimization}

As can be seen in Table \ref{tab:systems}, the transition from Titan to
Sierra/Summit is more than a simple increase in speeds and feeds: the
node density has also increased dramatically, with the result that
there are proportionally less CPU resources to manage the GPUs.  In
order to minimize the CPU overhead when running multi-GPU stencil
code, for communication within the node we utilize the CUDA IPC
interface which faciliates direct communication between GPUs within
the node.  Doing so has significant benefits: the NVLink connections
between GPUs in the node can be used optimally without being routed
though CPU memory maximizing the exchange bandwidth between GPUs in
the node.  Doing so also maximizes the inter-node performance since it
removes any contention for the inter-node traffic to the NIC through
the CPUs.  Finally, by utilizing direct CUDA DMA copies without MPI
synchronization, we avoid the need to synchronize the GPU and CPU for
managing the intra-node GPU transfers.  This minimizes the CPU
overhead: this is critical on dense-node systems where CPU resources
are more contended.  We note that while GPU-aware MPI does faciliate
the direct transfer between GPUs in a node, we are still forced to
synchronize the CPU and GPU.

The final step in this optimization is to utilize GPU Direct RDMA for
inter-node transfers, i.e., direct transfers between GPUs and NIC for
optimal routing of inter-node communication, further reducing CPU
overheads and maximizing inter-node scaling.  However, at the time of
submission the Sierra and Summit systems did not support this,
limiting our multi-node capability and scaling.

\noindent{\bf Communication Autotuning}
 
In this work, we relied heavily on improvements made in the QUDA
autotuner to improve the multi-node scalability, specifically
extending it to include the concept of {\it communication-policy}
tuning to pick the optimum communication approach for a given problem,
at a given node count on a given target machine.  While our use of
backfilling and \texttt{mpi\_jm} allows us to exploit the loop
parallelism of the problem (see below) it is still vital to maximize the
performance from exploiting the inherent data parallelism of each
linear solve since we will in general need a minimum number of GPUs
for a given calculation due to memory overheads, and moreover, the
outer loop over which we can parallelize, while large, is finite.

When deploying multi-process stencil computations on an MPI+GPU
system, there are a variety of options in how to coordinate the GPU
computation with the MPI communication.

\begin{itemize}
\item Utilize GPU DMA engines for transferring halo buffers to and from CPU and using regular MPI on the CPU 
\item Utilize zero-copy reads (writes) to (from) CPU memory for MPI sends (receives) on the CPU
\item Utilize GPU Direct RDMA for direct transfers between the GPU and NIC
\item Whether to wait for all communication to finish and launch a single halo update kernel for all dimensions (less latency) or use fine-grained communication that can achieve better overlap
\end{itemize}

The optimal approach will depend on many factors, including:
\begin{itemize}
\item Whether the NIC and GPU are on the same PCIe bus
\item DMA efficiency of the PCIe switch between the NIC and GPU
\item GPU architecture generation (pre-Pascal GPUs were only able to saturate EDR NIC bandwidth for NIC$\rightarrow$GPU transactions but not GPU$\rightarrow$NIC)   
\item Message size (smaller messages are clearly more latency bound, while DMA performance may diminish at large message size)
\item MPI and system software support for GPUs
\item Node density, e.g., how many GPUs are sharing the same NIC
\end{itemize}

Given this multi-dimensional parameter space, and given that it is
very machine specific, applying the autotuner to the
stencil-communication policy is very natural.  The end result is that
we achieve not only performance portability across GPU generations,
but also enables us to always use the optimum communication strategy
regardless of the machine topology and node count we are deployed on.

\noindent {\bf Management and Backfilling of Tasks}

Because of its statistical nature, a full Lattice QCD computation requires many intermediate-sized computational tasks that require different resources for different times.  Naively grouping even similar tasks into a single job creates the possibility of waste, as nodes can differ in performance, and it is often advantageous to group heterogeneous tasks together.  We found that naively bundling tasks---simply collecting and simultaneously launching HPC steps, and waiting for their completion---often caused a 20 to 25\% idling inefficiency.

A simple solution is to \emph{backfill} our computation, as a batch scheduler like SLURM or MOAB does.
This allows us to take advantage of the otherwise wasted idle time while still occupying easier-to-schedule large allocations.
As a simple proof of concept we wrote \texttt{METAQ} \cite{Berkowitz:2017vcp,Berkowitz:2017xna}, a set of shell scripts that forms a middle layer between the batch scheduler and the user's computational job scripts.  This simple software allowed us to recover an enormous fraction of our wasted time, effectively providing an across-the-board 25\% speed-up.

Due to its simplicity, \texttt{METAQ} is relatively hardware-agnostic and cannot, for example, guarantee that the nodes assigned to any task are near one another.  Furthermore, use of \texttt{METAQ} to manage many jobs requires a separate invocation of \texttt{mpirun} (or equivalent) for each task which can become taxing on the service nodes.

As we move towards the exascale era, where many applications in addition to lattice QCD, will require the computation of thousands, hundreds-of-thousands or millions of separate tasks with different resource requirements on heterogeneous architectures, it is critical to develop user-friendly, lightweight and efficient job-management tools that will enable the full utilization of these valuable machines.

To this end, we have developed \texttt{mpi\_jm}, a small library that, with just a few lines of added code, can be linked to in any application, and allows the executable to be launched with tight binding to the hardware.  Closeness to the hardware combined with a high-level python interface for resource management makes for a potent and intelligent backfilling tool.

Executables linked to \texttt{mpi\_jm} can be launched independently or under the management of the \texttt{mpi\_jm} scheduler.  That the same binary can function in both ways makes software maintenance and debugging easy.  

The job manager \texttt{mpi\_jm} is started as a collection of mpirun launches of a single-node manager process per node on groups of nodes (for example 32 or 128 nodes) that we call lumps.   The first lump also starts a scheduler process and the remaining lumps connect to the scheduler after they initialize.  The connection process uses the DPM (Dynamic Process Management) features of MPI 3.1.   At the moment the only suitable MPI implementations that support the required DPM features are MPICH~\cite{mpich} and MVAPICH2~\cite{mvapich}, which is built on an MPICH base.

Interconnect performance is controlled by further subdividing lumps into blocks.
Blocks are configured to have a number of nodes that is a multiple of the largest jobs being run, say 4 or 8 nodes while a lump might be 64 or 128 nodes.    The member nodes of a block are chosen to be close together for high performance communications.    In \texttt{METAQ} we see that as jobs of different sizes complete and start again that the available nodes became fragmented, impacting performance.    In \texttt{mpi\_jm} the block boundaries prevent fragmentation and keep high bandwidth communications local.

The workload of jobs is defined via a python interface to an underlying C++ based scheduler.   The scheduler also reads a python based description of the nodes, detailing the memory, cores, slots, and GPUs.   This information is used to automatically assign ranks to appropriate nodes and slots on those nodes to efficiently access the GPUs.    The scheduler has a detailed map of resources consumed by running jobs and finds good matches between ready jobs and open resources.
Next a job start message is sent to a block with the open resources and it in turn uses \texttt{MPI\_Comm\_spawn\_multiple} to start the job on the assigned resources.    We made an attempt to implement a firewall against the failure of the individual job by disconnecting the intracommunicator resulting from the spawn but found that a call to \texttt{MPI\_Abort} in a disconnected job still brings the entire lump down (in violation of the MPI standard), but fortunately not the entire system.   This led us to use relatively small lump sizes on new systems that may be suffering from pre-acceptance issues.

Each launch of a lump is on a bounded number of nodes and hence does not suffer from the common non-linear startup cost for large sets of nodes.   Any lumps that fail to start due to a node with damaged connectivity or file system problems don't connect and are ignored.   Even on thousands of nodes this partitioned startup process is very fast, taking only a couple of minutes.    Once the lumps have started and connected, the scheduler begins distributing jobs to the lumps where they are started in parallel.   
On Sierra, we were able to bring a 4224 node job up and running in 3-5 minutes, taking advantage of the lumps and MVAPICH2.
In less than one minute, all lumps were connected and within five minutes, nearly all nodes were performing real work.

A major advantage of \texttt{mpi\_jm} is that users can easily and safely run multiple executables on the same node.  On new machines with dense nodes, when the intensive numerical work is offloaded to the GPUs, in many applications the CPUs sit idle or have very little work.  With this scheduler, users can easily cut nodes into smaller pieces, or overlay GPU-capable and CPU-only tasks on the same nodes, minimizing idle cycles.
This creates the possibility of taking advantage of an often-neglected additional resource, so that CPU-only tasks become effectively free.

We installed \texttt{mpi\_jm} in the communications layer of the community Lattice QCD software, \texttt{QMP}\cite{githubQMP}, so that all LQCD applications can take leverage this management easily.
We have used it to produce results on Sierra, ensuring we can do jobs as small as one computational task and scaling up to $> 3000$ nodes.
On Sierra, with a large scale job, we have successfully overlapped numerically-intensive tasks that require GPUs with simpler CPU-only tasks, paving the way to hide a nontrivial computational cost.

\section{\label{sec:perf_meas_method} Performance Measurement Method}

\noindent {\bf Overview}

\begin{figure}[!t]
  \centering
  {\includegraphics[width=1.0\columnwidth,angle=0]{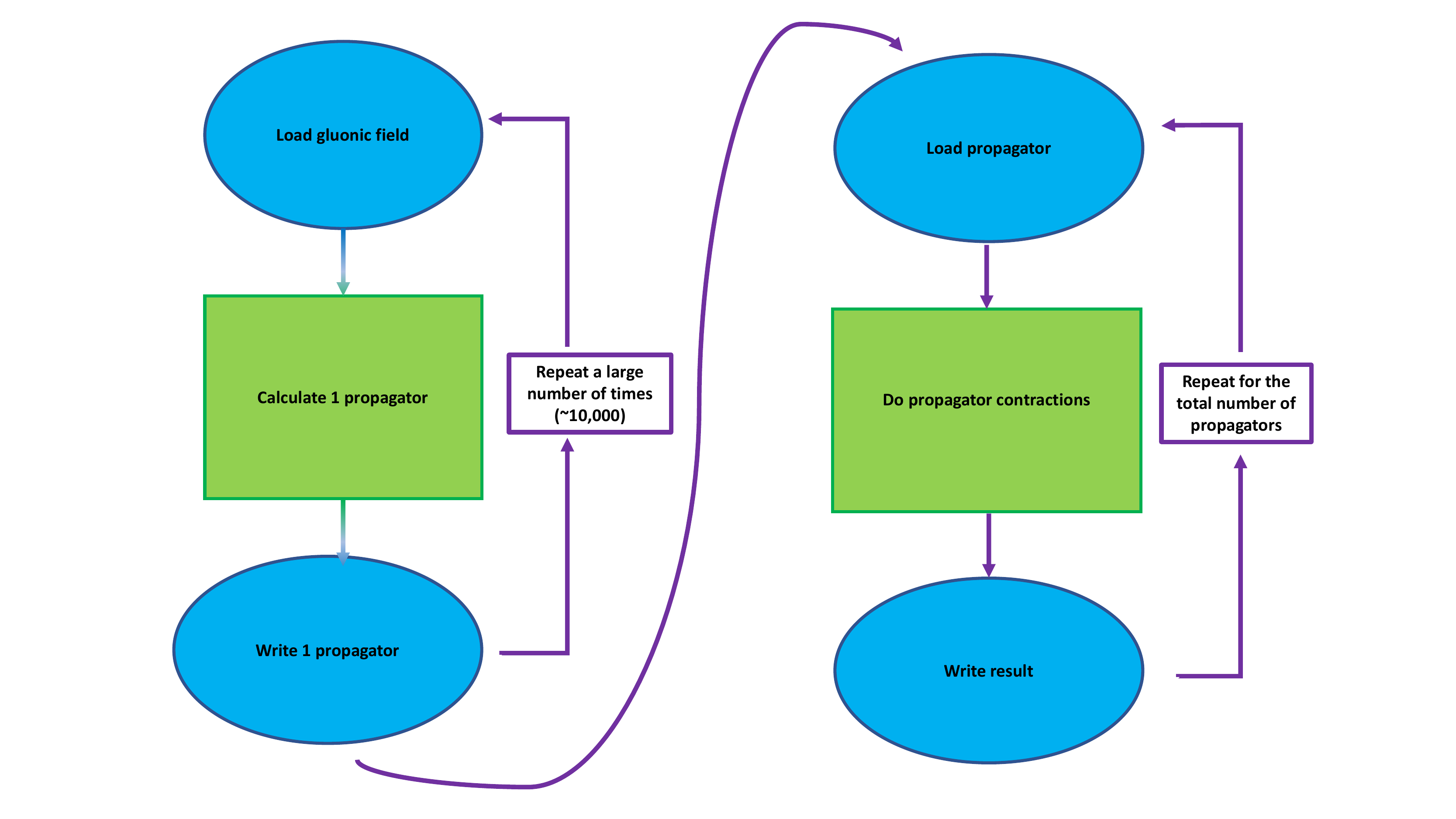}}
  \caption{The workflow of our application. I/O (blue ovals) is performed when we read a gluonic field configuration, write and read propagators, and write final results.  Computation (green boxes) entails solving the Dirac equation for propagators and performing tensor contractions with those propagators.}
  \label{fig:workflow}
\end{figure}

Because LQCD is a Monte Carlo method, for each lattice size we have a
large ensemble of \emph{gluonic field configurations} which dictate
how the quarks move.  The workflow is shown schematically in in
Figure~\ref{fig:workflow}.  On each configuration we solve the Dirac
equation for large number of numerically expensive quark propagators.
These propagators are ingredients in tensor contractions that yields
the numerical data of interest; a single tensor contraction depends on
more than one propagator.  To control our systematic effects, such as
from spacetime discretization, we use many ensembles, varying the
lattice sizes and other parameters.  Each ensemble requires solves for
propagators and tensor contractions.  For each ensemble, the
propagator solves consume about 97\% of the execution time, while
tensor contraction consumes about 3\%.

\noindent {\bf Linear Solver Performance}

When reporting the linear solver performance, we utilize three
performance metrics: sustained floating-point operating per second,
sustained effective memory bandwidth, and sustained percent of
floating-point peak.  While the flops rate is a relevant measure for
machine comparisons, as a relative measure of machine efficiency,
sustained bandwidth is the most relevant since the limiting
performance of a stencil computation should in almost all cases be
bandwidth.

For reporting flops, we simply add up the necessary number of
operations required in the stencil application and the auxiliary
BLAS-1 operations required in the CG linear solver.  This is
straightforward to do, since the stencil is uniform and acts on a
regular grid.  For the red-black-preconditioned Domain-wall stencil
used in this work, there are between 10,000-12,000 floating point
operations per five-dimensional lattice point (using conventions
consistent in the LQCD domain).  In comparison, the
BLAS-1 operations have 50-100 flops per lattice site, i.e., they are
extremely bandwidth bound.  When utilizing 16-bit precision, the
overall arithmetic intensity of the CG solver varies from 1.8-1.9,
thus we can convert our flops number into effective sustained
bandwidth by dividing by the arithmetic intensity.

To correctly report percent of peak, we cannot directly utilize the
standard flops rate, since doing so would exclude the fact that not all arithmetic
operations can be issued as fused-multiply-add instructions, as well
as the fact that all reductions are done in double precision.  Thus
when reporting percent of peak, we take into account these factors,
which corresponds to approximately a \(1.675\times\) scaling on the raw
solver flops rate, and report the peak as a percentage of
single-precision peak flops rate.
 
For timing these GPU performance rates we utilize timers on the CPU,
ensuring to synchronicity with the GPU when start and end times are
recorded.

\noindent{\bf CPU / GPU concurrency}

The contractions are CPU-only tasks and due to our job manager,
\texttt{mpi\_jm}, we concurrently run them on the same nodes as the
propagators. This completely amortizes their cost. With
\texttt{mpi\_jm} we carve out sections of nodes to compute propagators
and do the same for contractions of previous propagators that have
already been written to disk. For reading and writing we have
interfaced with the parallel hdf5 library \cite{Kurth:2015mqa}. I/O
takes about 0.5 \% of our total application time, even after
accounting for this interweaving of CPU/GPU jobs, so we do not include
it in our total application budget.

\noindent{\bf Systems Tested}

We have utilized the newly-installed CORAL systems, Sierra and Summit,
at Livermore and Oak Ridge National Laboratories, respectively, in
this work.  The specifications of these systems are summarized in
Table \ref{tab:systems}, where we also include the prior Titan system,
and the pre-CORAL development system at LLNL, Ray, which can be
considered a stepping stone between Titan and Sierra/Summit since it
utilizes the intermediate Pascal GPU architecture.  We also show on
this table the various compiler and library versions used on each machine.

\begin{table}[h]
\centering
\begin{tabular}{|c|c|c|c|c|}
\hline
Attribute                    & Titan & Ray & Sierra & Summit \\ \hline
\hline
nodes & 18,688 & 54 & \(\sim\)4200 &  \(\sim\)4600 \\ \hline
GPUs / node & 1 & 4 & 4 &  6  \\ \hline
CPU & AMD & IBM & IBM & IBM \\
 & Opteron & POWER8 & POWER9 & POWER9 \\ \hline  
GPU & NVIDIA & NVIDIA & NVIDIA & NVIDIA \\
& K20X & P100 & V100 & V100 \\\hline  
\hspace{-2em}FP32 TFLOPS\hspace{-2em} & 4 & 44 & 60 & 90 \\ \  
/ node & &  &&    \\ \hline
GPU bw & 250 & 2880 & 3600 & 5400 \\ \  
/ node GB/s & &  &&    \\ \hline
CPU-GPU & 6 & 20 & 75 & 50 \\
bw GB/s&  &  &  &  \\ \hline
Interconnect & Cray Gemini& Mellanox & Mellanox & Mellanox \\
& (\(\sim\) 8 GB/s) & IB 2xEDR & IB 2xEDR & IB 2xEDR \\  \hline 
GCC & 4.9.3 & 4.9.3 & 4.9.3 & 4.8.5 \\ \hline
MPI &  Cray MPICH & Spectrum & \hspace{-2em}MVAPICH2\hspace{-2em} & Spectrum \\
& 7.6.3 & 2017.04.03 & 2.3 & \hspace{-1em}2018.01.10 \hspace{-1em} \\ \hline
CUDA & 7.5.18 & 9.0.176 & 9.2.148 & 9.1.85 \\ \
toolkit&  &  &  & \\ \hline
\end{tabular}
\caption{Comparison of the systems used in this study}
\label{tab:systems}
\end{table}

On all machines we used the same physics application software.  We
summarize the various packages and libraries used in table
\ref{tab:sw}, together with specific version used, and the
repositories for these packages. 

\begin{table}[h]
\centering
\begin{tabular}{|c|c|c||}
\hline
Name                     & commit id & repository \\ \hline
\hline
Lalibe & N/A & https://github.com/callat-qcd/lalibe \\ \hline
Chroma & 72a47bd & https://github.com/JeffersonLab/chroma \\ \hline
QUDA & 6d7f74b &  https://github.com/lattice/quda \\ \hline
QDP++ & 5b711236 & https://github.com/azrael417/qdpxx.git \\ \hline
QMP & d29f3f8 & https://github.com/callat-qcd/qmp \\ \hline
mpi\_jm & a4722f5 & https://github.com/kenmcelvain/mpi\_jm \\ \hline
\end{tabular}
\caption{Application software used in this study.
QDP++ was forked from \url{https://github.com/usqcd-software/qdpxx} and QMP was forked from \url{https://github.com/usqcd-software/qmp}; our changes will be merged soon.}
\label{tab:sw}
\end{table}

\section{\label{sec:perf_results} Performance Results}

\begin{figure}[!t]
  \centering
  \subfigure[]{\includegraphics[width=.80\columnwidth,angle=0]{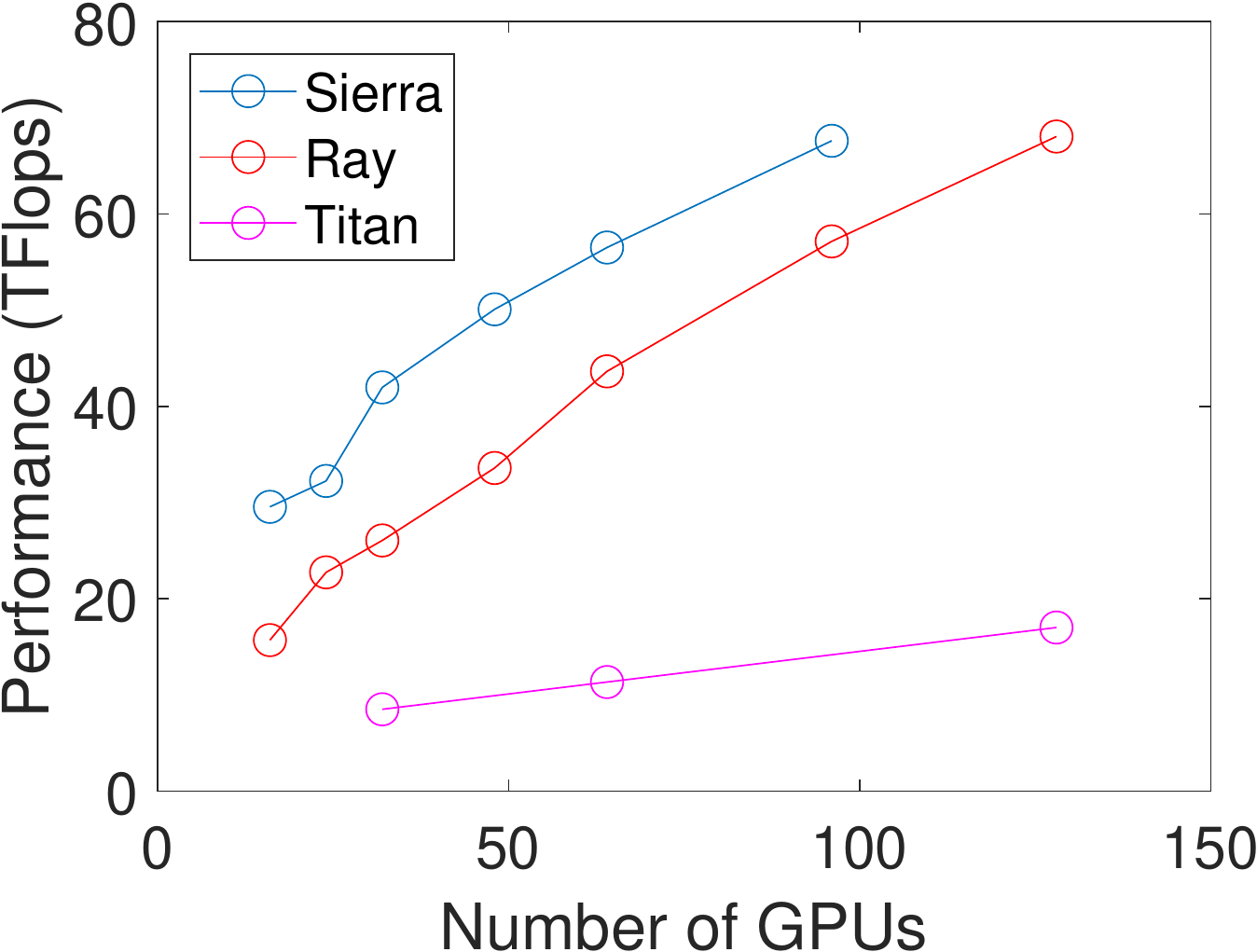}} \\
  \subfigure[]{\includegraphics[width=.80\columnwidth,angle=0]{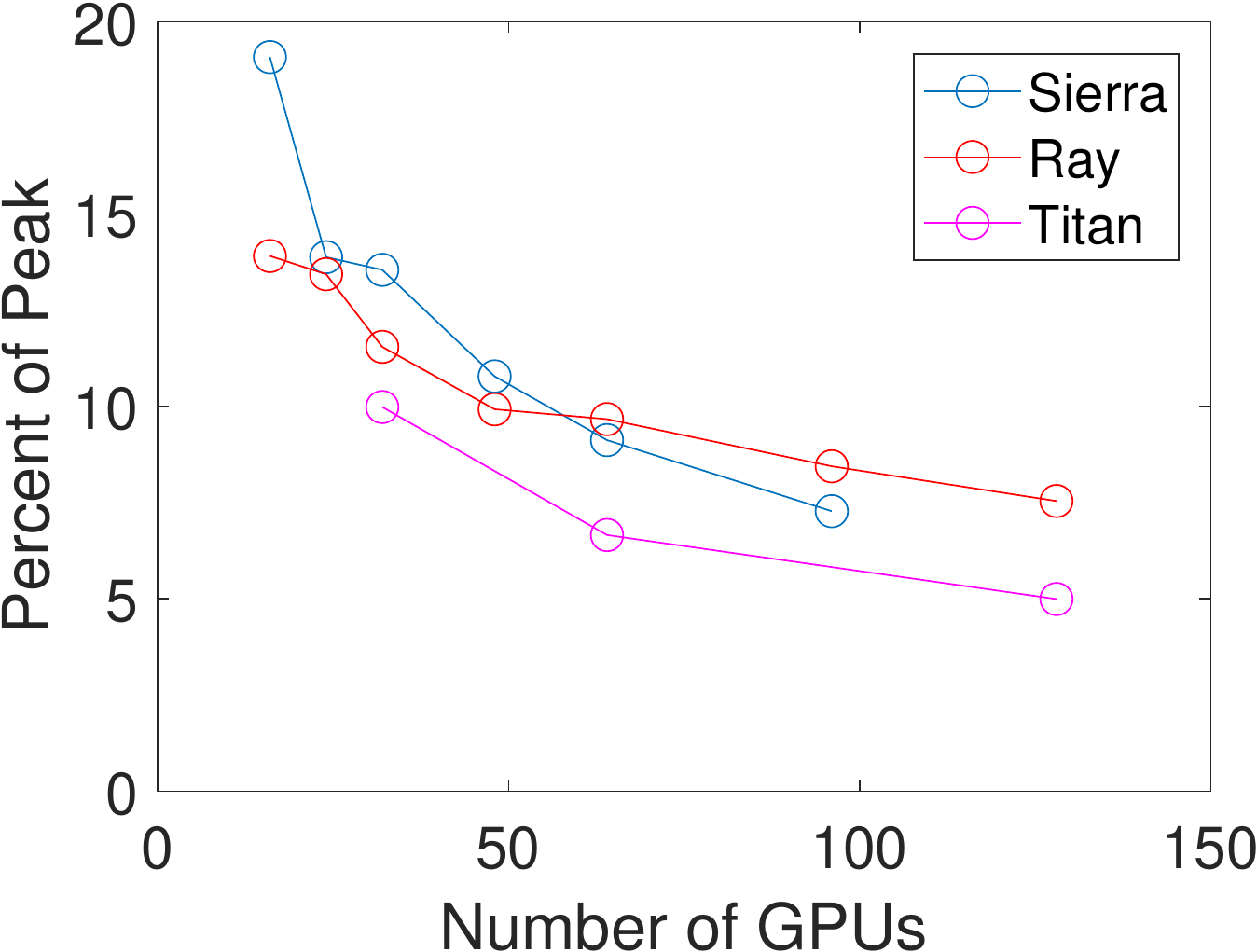}} \\
  \subfigure[]{\includegraphics[width=.80\columnwidth,angle=0]{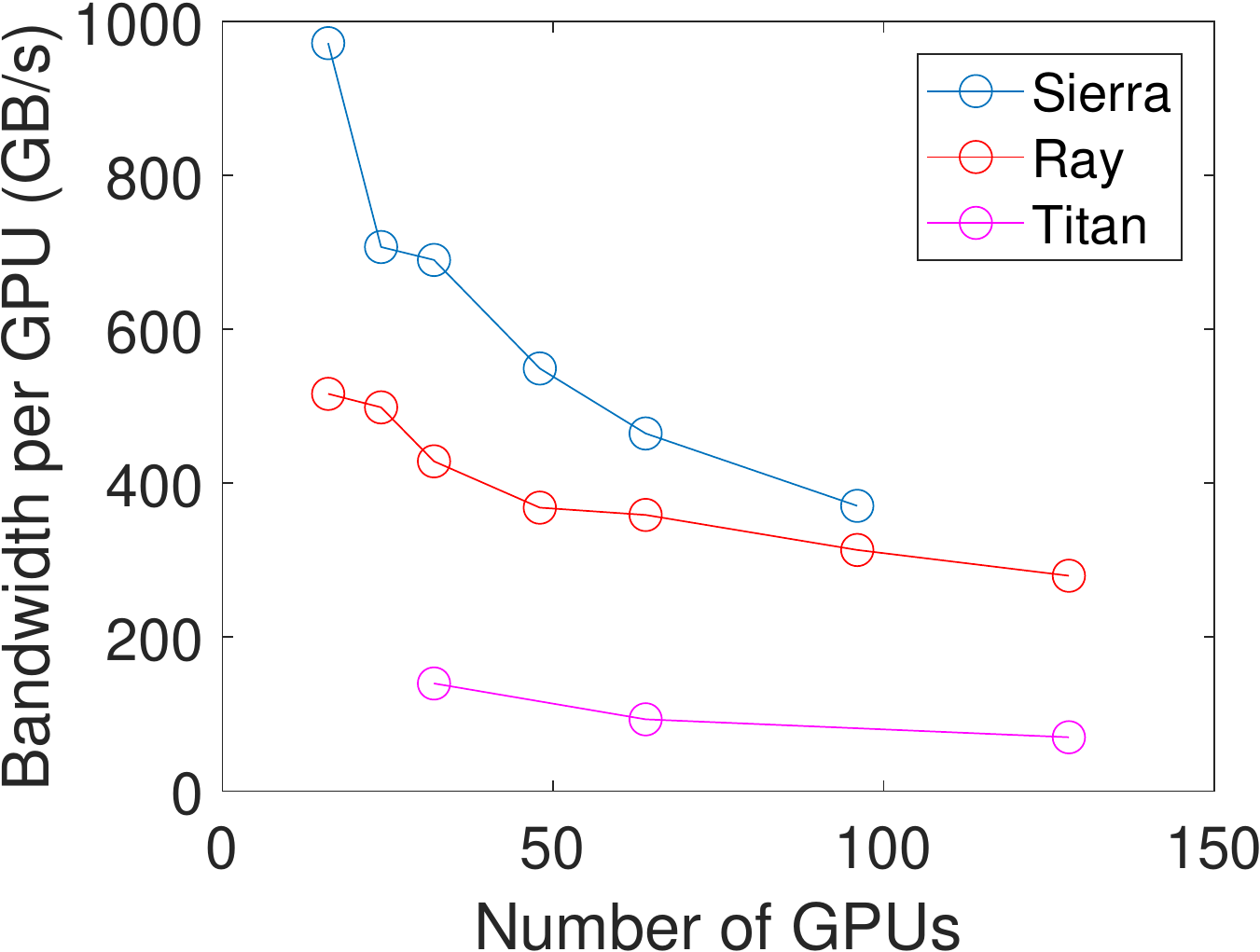}} \\
  \caption{ Performance of QUDA's CG algorithm on Titan, Ray, and Sierra are shown above. We report performance in both TeraFlops in figure (a) and percent of peak in figure (b). All three machines are solving for a propagator on a  $48^3 \times 64$ lattice. In figure (c) we give bandwidth performance, which we obtain by converting raw performance per GPU to GB/s by using an arithmetic intensity of 1.9.
}
  \label{fig:strongscale_subgroup}
\end{figure}

In a typical workflow, we first perform strong-scaling tests over a
single propagator calculation to determine the optimal number of nodes
to carve out using \texttt{mpi\_jm}.  We do so in figure
\ref{fig:strongscale_subgroup}, where we show the strong scaling for
the $48^3 \times 64$ volume, comparing the performance over three
different generations of GPU architecture.  We observe that the
expected behavior of increased performance with successive GPU
generation is obtained, but what is more interesting is that we see
the maximum percent of peak performance achieved increases with
successive GPU architectures.  If we convert the aggregate sustained
performance in TFLOPS to bandwidth per GPU at the point of peak
efficiency, e.g., lowest GPU count, we see values of 139 GB/s,
516 GB/s, and 975 GB/s for Titan, Ray and Sierra, respectively.  This
we attribute to the improved cache structure of successive GPUs which
has seen a steady increase to both the L1 and L2 cache available per
thread: this has the effect of amplifying the effective bandwidth.

\begin{figure}[!t]
  \centering
  {\includegraphics[width=.80\columnwidth,angle=0]{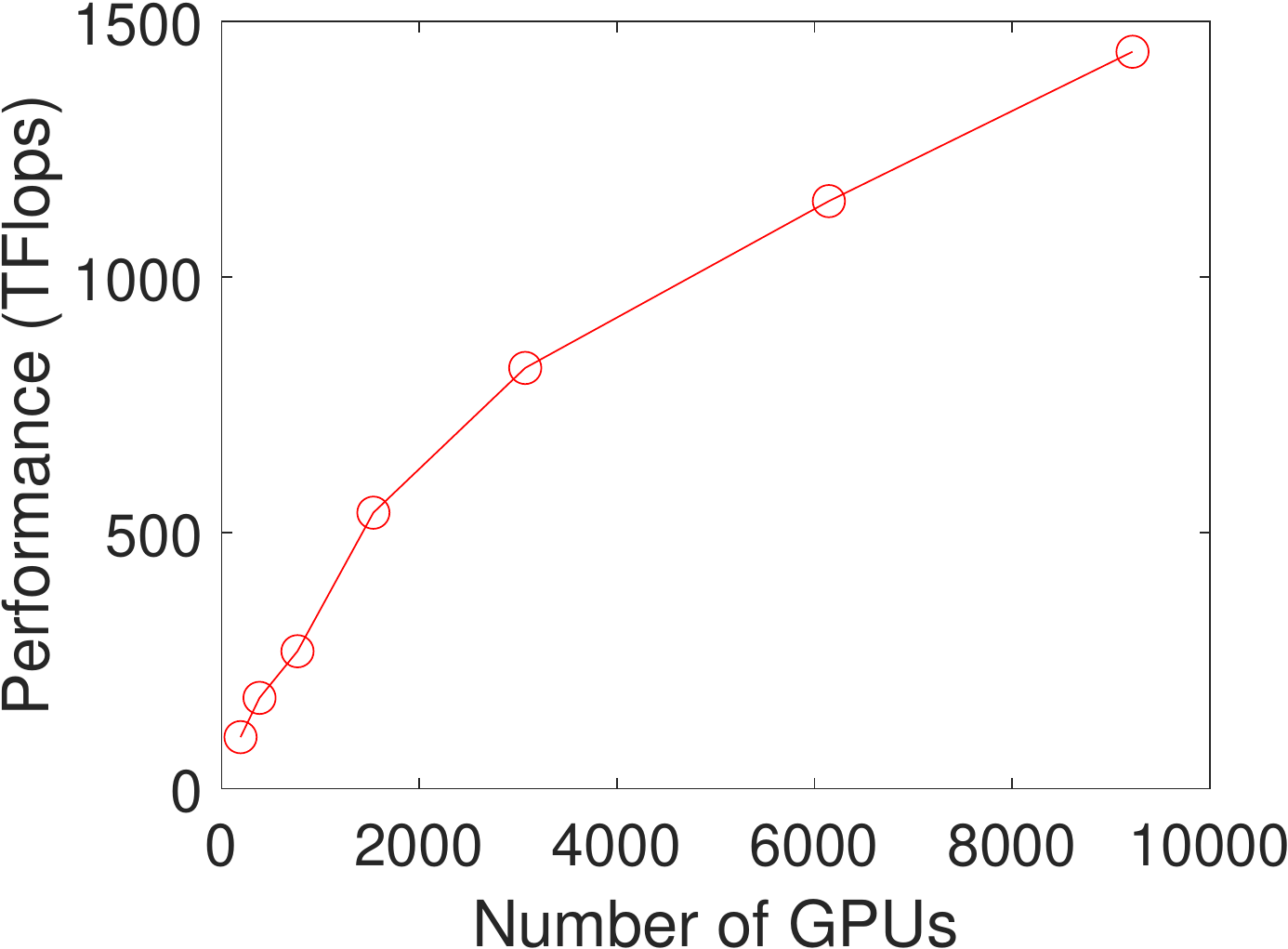}}
  \caption{Strong scaling performance of our propagator routine on Summit with a single $96^3 \times 144$ lattice.
}
  \label{fig:Summitperf}
\end{figure}

In figure \ref{fig:Summitperf} as a next-generation proof-of-concept
problem, we show strong scaling on Summit for the much larger $96^3\times144$
volume.  As expected, a much larger problem size allows us to strong
scale to a significant fraction of the machine, and the sustained
solver performance approaches 1.5 PFLOPS.  While this is may seem
significant, we note that there is a large drop in solver efficient
past \( \sim 2000\)
GPUs, with dramatically reduced solver efficiency demonstrating that
we cannot rely on simple data-parallel strong scaling alone in order
to saturate large machines.

\begin{figure}[!t]
  \centering
{\includegraphics[width=.80\columnwidth,angle=0]{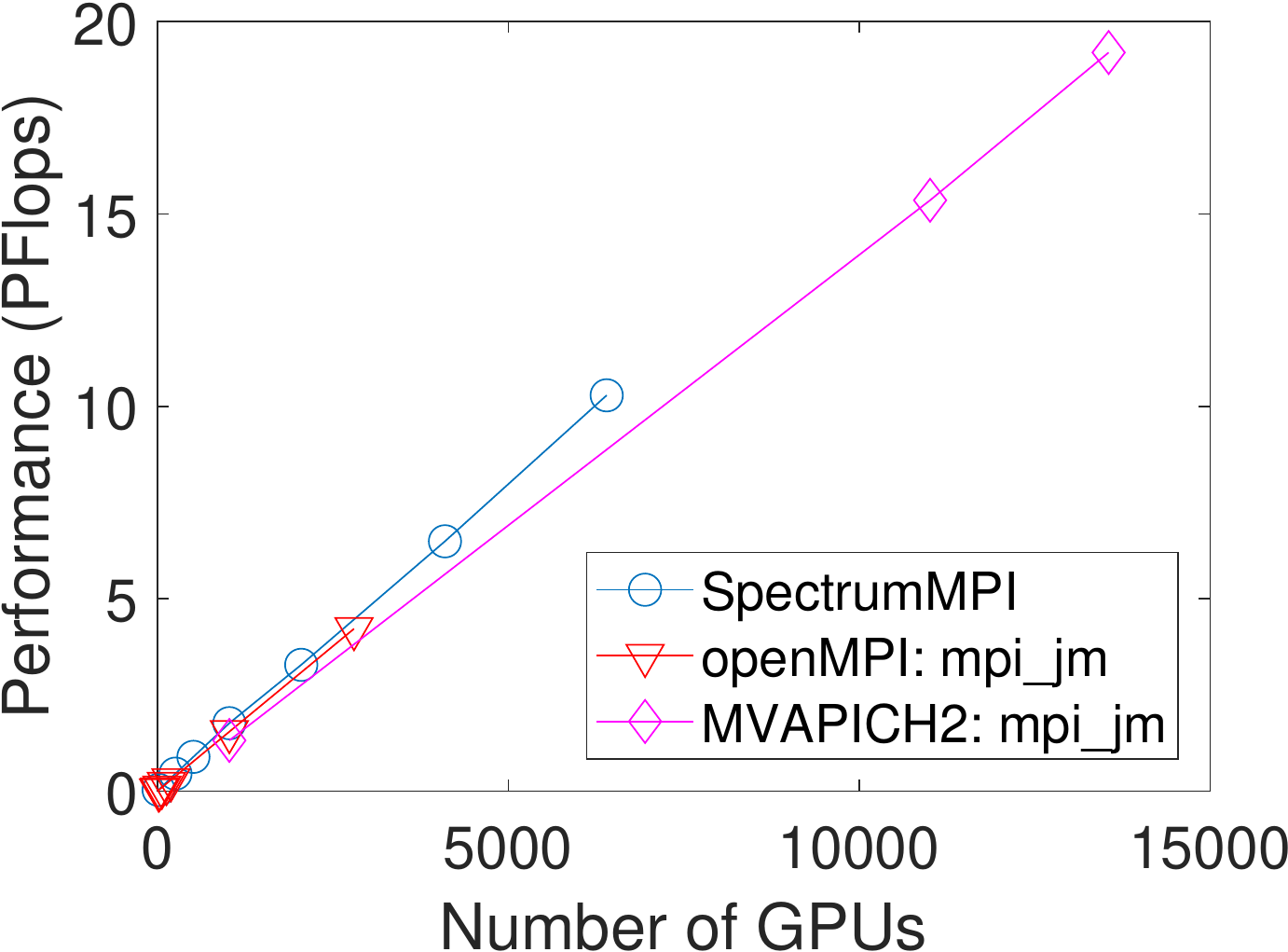}}
\caption{\label{fig:strong-weak_scale_sierra}
The sustained performance in PFlops as the number of propagator calculations is increased.
These calculations are done in groups of 4 nodes (16 GPUs) on Sierra, with a $48^3 \times 64$ lattice.
The SpectrumMPI jobs were submitted as individual jobs to the scheduler, with 400 jobs submitted for the most performant run.
The openMPI: mpi\_jm jobs were run (in April) in up to 7 blocks of 100 nodes, where each block was a separate mpi\_jm job managing the sub-tasks.
The MVAPICH2: mpi\_jm jobs were submitted as a single job to the scheduler with all nodes managed by a single mpi\_jm task.
}
\end{figure}

Instead, we utilize \texttt{mpi\_jm} to
determine dependencies and optimally distribute simultaneous
propagators and contractions. This leads to excellent strong scaling
in the number of groups that \texttt{mpi\_jm} manages and allows us to
sustain the highest performance from figure
\ref{fig:strongscale_subgroup}. Hardware cross interactions are
negligible and from figure \ref{fig:strong-weak_scale_sierra} we see
that weak scaling is nearly perfect.


In figure \ref{fig:strong-weak_scale_sierra} we compare weak scaling
by increasing the number individual propagator solves with the number
of GPUs on Sierra. We also compare this to deployment of solves with a
single call to \texttt{mpi\_jm}.  
This suffers from a slight performance hit because we used
MVAPICH2 to obtain the required dynamic process management features,
but have not yet fully tuned its settings for Sierra.

\begin{figure}[!t]
  \centering
{\includegraphics[width=.80\columnwidth,angle=0]{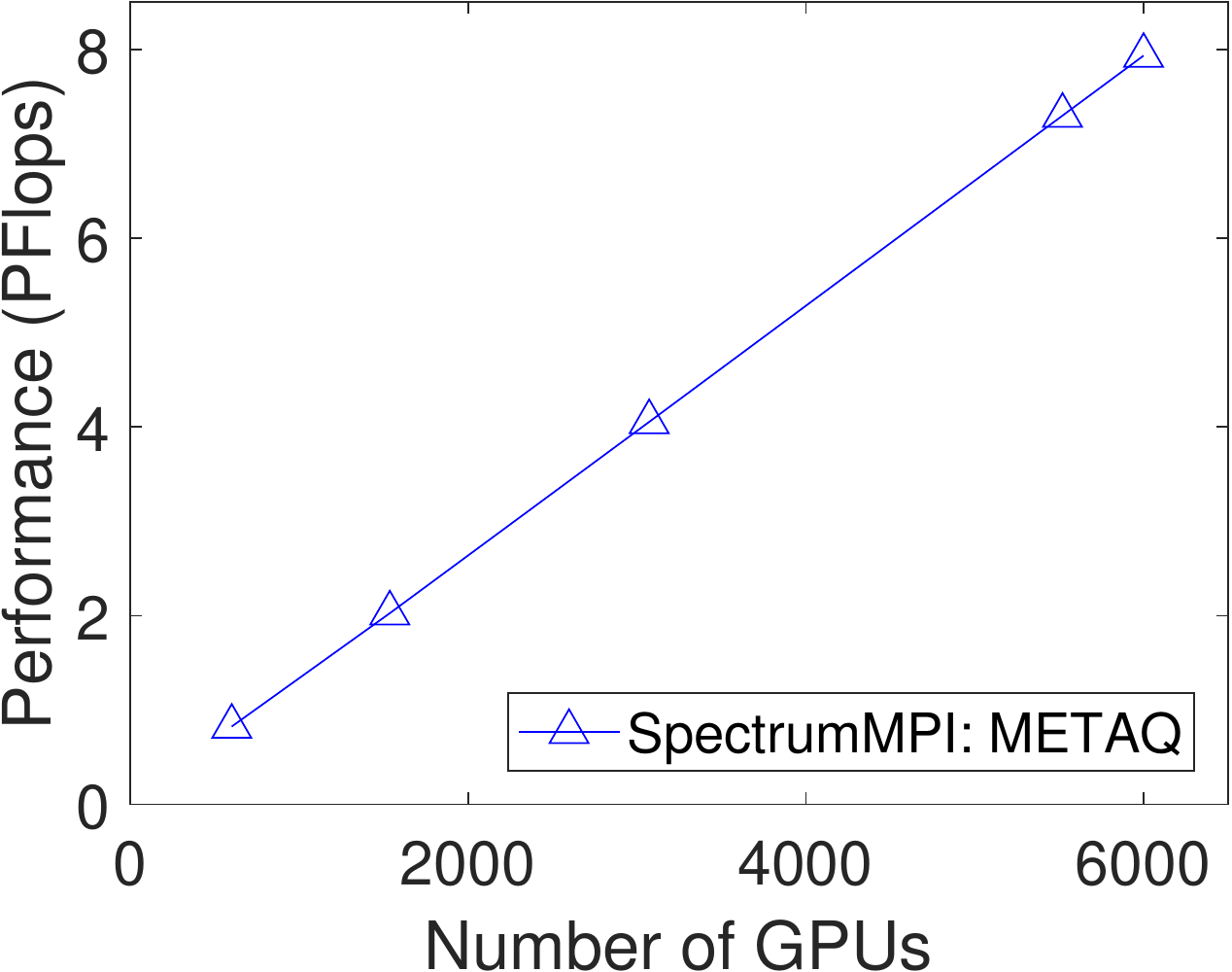}}
\caption{\label{fig:metaq_summit}
The sustained performance in PFlops as the number of propagator calculations is increased.
These calculations are done in groups of 4 nodes (24 GPUs) on Summit, with a $64^3 \times 96$ lattice.
The jobs were managed with a single instance of METAQ that used \texttt{jsrun} to start each individual propagator task within a given allocation.
}
\end{figure}

Figure \ref{fig:metaq_summit} shows weak scaling on Summit with \texttt{METAQ}. Once again, our job management achieves perfect weak scaling. We would like to note, that with mpi\_jm, we are able to place different jobs on different resources on a node.  This has allowed us, for example, to simultaneously run independent GPU-intensive and CPU-only jobs on the same nodes.  We have demonstrated this at scale on Sierra.  This can be particularly advantageous for Summit and problems which do not have powers of three that nicely map onto the 6 GPUs per node.  As a simple example, a set of three jobs that require 16 GPUs each can nicely be placed on 8 Summit nodes (48 GPUs).  The first and second jobs can occupy GPUs 1,2,4,5 on nodes 1-4 and 5-8, while the third job can be placed on GPUs 3,6 on all 8 nodes.  We have demonstrated this job placement on Summit with our workload.  While the jobs that occupy 2 GPUs per node suffer a performance degredation, this can be largely mitigated by the backfilling capability of mpi\_jm, such that all GPUs (and CPUs) on the node will remain occupied with new jobs as the faster ones complete and free their resources.


\begin{figure}[!t]
  \centering
{\includegraphics[width=.80\columnwidth,angle=0]{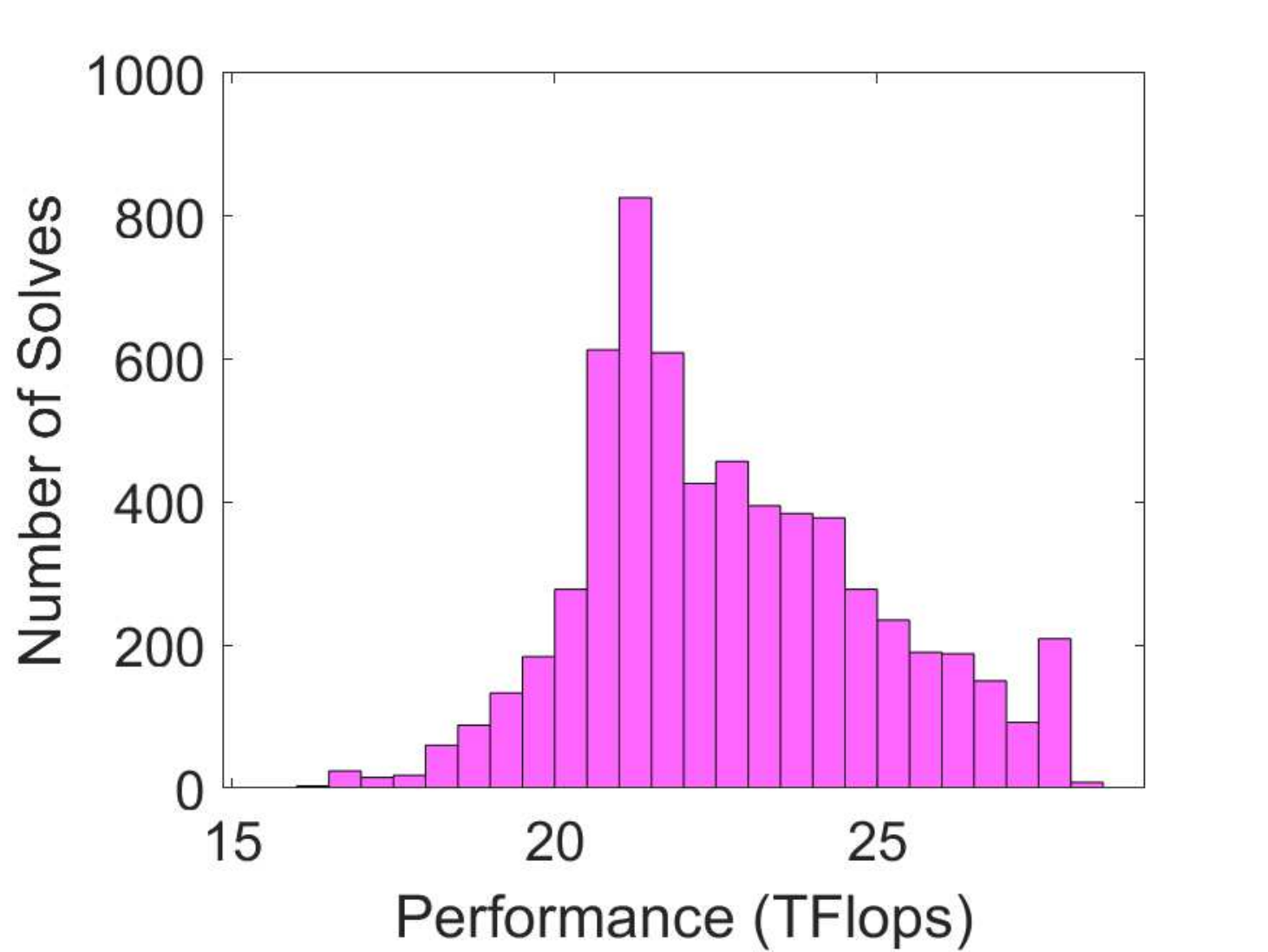}}
\caption{\label{fig:hist_sierra}
Histogram of the solver performance when running on Sierra. This result is from our largest run with 13500 GPUs, using \texttt{mpi\_jm} with MVAPICH2.
}
\end{figure}

To compute sustained performance of our application we account for the performance of all tasks, propagator generation, contractions, and I/O. Propagators take 96.5 \% of the computation, contractions take 3 \%, and I/O 0.5 \%. I/O is completely negligible and while our contractions account for only a small fraction, by interleaving them on the CPUs of nodes that have GPUs running propagators, their cost is brought to zero. This leaves only propagator solves, resulting in a sustained performance of 20\% on the minimal number of nodes.  However, when we scaled to large fractions of the machine, we relied on our job manager which, as mentioned, needed MVAPICH2.  When we run a small job with MVAPICH2 we get a sustained performance equivalent to that of SpectrumMPI. When we learn how to fully utilize MVAPICH2, we anticipate bringing the sustained performance at scale from 15\% to 20\%, which is the right edge of figure \ref{fig:hist_sierra}.

The combination of the improvements to \texttt{QUDA} (the GPU lattice QCD library) and the development of \texttt{mpi\_jm}, have allowed us to scale our workload of 4-node jobs to 3388 Sierra nodes under a single job submitted to the scheduler, with a startup time of about 5 minutes for the full set of nodes.  With small improvements to \texttt{mpi\_jm}, we will be able to launch a workload to the full scale of Sierra or Summit without the loss of any good nodes.
At present, we have achieved a peak sustained performance on Sierra of nearly 20 PFlops, which amounts to 15\% of peak performance.  This is the top performance in a single job-submission that has been achieved for lattice QCD to date.
Stated differently, the machine-to-machine speed up of Sierra and Summit over Titan, for our research program, is a factor of approximately 12 and 15 respectively.  These CORAL machines are truly disruptive and we look forward to using them as extensively as possible.


\section*{Acknowledgements}

We are indebted to the Livermore Computing Center for access to Sierra and help getting set up there.
In particular, our contacts John Gyllenhal, Adam Bertsch and Py Watson were very helpful and responsive.
Our other Livermore colleagues who also applied for an award were very cooperative and collaborative, keeping us appraised of the state of the machine, difficulties they encountered, and workarounds.  In particular, Jim Glosli, Tomas Oppelstrup, and especially Tom Scogland were very generous with their time and concern.
At the Oak Ridge Leadership Computing Facility, Jack Wells provided excellent help, advice, and encouragement.

An award of computer time was provided by the Innovative and Novel Computational Impact on Theory and Experiment (INCITE) program to CalLat (2016) as well as the Lawrence Livermore National Laboratory (LLNL) Multiprogrammatic and Institutional Computing program through a Tier 1 Grand Challenge award. This research used the NVIDIA GPU-accelerated Titan, Summit-Dev, and Summit supercomputers at the Oak Ridge Leadership Computing Facility at the Oak Ridge National Laboratory, which is supported by the Office of Science of the U.S. Department of Energy under Contract No. DE-AC05-00OR22725, and the NVIDIA GPU-accelerated Surface, Ray, and Sierra supercomputers LLNL.  This work was performed under the auspices of the U.S. Department of Energy by LLNL under Contract No. DE-AC52-07NA27344 and under contract DE-AC02-05CH11231, which the Regents of the University of California manage and operate Lawrence Berkeley National Laboratory and the National Energy Research Scientific Computing Center.
The USQCD (QMP, QDPXX, Chroma) and CalLat (HDF5 in QDPXX, METAQ) software used and developed for this work was supported in part by the U.S. Department of Energy, Office of Science, Office of Advanced Scientific Computing Research and Offices of Nuclear and High Energy Physics, Scientific Discovery through Advanced Computing (SciDAC) program.

\bibliographystyle{IEEEtran}
\bibliography{GBbib}

\end{document}